\documentclass[twocolumn,english,aps,pra,showpacs,tightenlines]{revtex4}
\usepackage{amsmath}
\usepackage{graphicx}

\begin{document}

\title{Adiabatic quantum state transfer in non-uniform triple-quantum-dot
system}
\author{Bing~Chen, Wei~Fan, and Yan~Xu}
\affiliation{College of Science, Shandong University of Science and Technology, Qingdao
266510, China}

\begin{abstract}
We introduce an adiabatic quantum state transfer scheme in a non-uniform
coupled triple-quantum-dot system. By adiabatically varying the external
gate voltage applied on the sender and receiver, the electron can be transferred between
them with high fidelity. By numerically solving the master equation for a system with \emph{always-on} interaction, it is indicated that the transfer fidelity depends on the ration between the peak voltage and the maximum coupling constants. The effect of coupling mismatch on the transfer fidelity is also investigated and it is shown that there is a relatively
large tolerance range to permit high fidelity quantum state transfer.
\end{abstract}

\maketitle

\section{Introduction}

In quantum information science, quantum state transfer (QST), as the name
suggests, refers to the transfer of an arbitrary quantum state from one
qubit to another. Recently, there are two major mechanisms for QST. The
first approaches are usually characterized by preparing the quantum channel
with an \textit{always-on} interaction where QST is equivalent to the time
evolution of the quantum state in data bus~\cite{Bose1,Song,Christandle1}.
However, these approaches require precise control of distance and timing.
Any deviation may leads to significant errors. The other approaches have
paid much attention to adiabatic passage for coherent QST in time-evolving
quantum systems. The most well known example of these is the so-called
Stimulated Raman Adiabatic Passage (STIRAP) technique, which is used to
produce a complete population transfer between two internal quantum states
of an atom or molecule~\cite{Shore}. Such methods are relatively insensitive
to gate errors and other external noises and do not require an accurate
control of the system parameters, thus can realize high-fidelity QST.

Due to the potential scalability and long decoherence times, the
applications of adiabatic passage have been widely investigated in
solid-state systems~\cite%
{Vitanov,TB,Eckert,Zhang,GT1,GT2,GT3,BEC1,BEC2,BEC4,DAS,McE}. Eckert et al.~%
\cite{Eckert} have introduced an implementation of the STIRAP in the
three-trap potential array. By coherently manipulating the trap separation
between each two traps, the neutral atoms can be transferred in the
millisecond range. Zhang et al.~\cite{Zhang} have describe a scheme for
using an all-electrical, adiabatic population transfer between two spatially
separated dots in a triple-quantum-dot (TQD) system by adiabatically
engineering external gate voltage. In ref.~\cite{GT1}, A. D. Greentree et
al. have described a method of coherent electronic transport through a
triple-well system by adiabatically following a particular energy eigenstate
of the system. By adiabatically modulating coherent tunneling rates between
nearest neighbor dots, it can realize a high fidelity transfer. This method
was termed Coherent Tunneling by Adiabatic Passage (CTAP) which was
demonstrated experimentally very recently via optical waveguide~\cite{Longhi}%
. Since then, adiabatic passage has also been used to transport quantum
information from a single sender to multiple receivers, which relates to a
quantum wire or fan-out~\cite{GT3}. Following a different perspective, there
have been several recent proposals to coherently manipulate BECs~\cite%
{BEC1,BEC2,BEC4} in triple-well potentials. Ref.~\cite{DAS} has analytically
derived the condition for coherent tunneling via adiabatic passage in a
triple-well system with negligible central-well population at all times
during the transfer.

In CTAP technique~\cite{GT1}, the basic idea is to use the existence of a
spatial dark state which is a coherent superposition state of two
\textquotedblleft distant\textquotedblright\ spatial trapping sites,
\begin{equation*}
\left\vert D_{0}\right\rangle =\cos \theta _{1}\left\vert L\right\rangle
+0\left\vert M\right\rangle -\sin \theta _{1}\left\vert R\right\rangle ,
\end{equation*}%
where the mixing angle $\theta _{1}$ is defined as $\tan \theta _{1}=\Omega
^{LM}/\Omega ^{MR}$ with $\Omega ^{LM}$ ($\Omega ^{MR}$) denoting the
tunneling rate between nearest-neighbor dots. By maintaining the system in
state $\left\vert D_{0}\right\rangle $ and adiabatically manipulating the
tunneling rates, it is possible to achieve coherent population transfer
from site $\left\vert L\right\rangle $ to $\left\vert R\right\rangle $\
without any probability being in the state $\left\vert M\right\rangle $. In
this paper we consider a different adiabatic protocol to achieve population
transfer between two spatially separated dots. We introduce a non-uniform
coupled triple-quantum-dot array which can be manipulated by external gate
voltage applied on the two external dots (sender and receiver). Through
maintaining the system in the ground state we show that the electron
initially in the left dot can be transferred to the right dot occupation
with high fidelity. Furthermore, we study in details the dynamic competition
between the adiabatic QST and the decoherence. There are two time scales $%
T_{1}$ and $T_{2}$ depicting such competition, where $T_{1}$ represents the
adiabatic time limited by the adiabatic conditions and $T_{2}$ represents
the decoherence time.

The paper is organized as follows. In Sec. II we setup the model and we
describe the adiabatic transfer of an electron between quantum dots. We also
derive a perturbative, analytical expression of fidelity. In Sec. III we
show numerical results that substantiate the analytical results. The last
section is the summary and discussion of this paper.

\section{Model setup}

\begin{figure}[tbp]
\includegraphics[ bb=188 384 450 524, width=6 cm, clip]{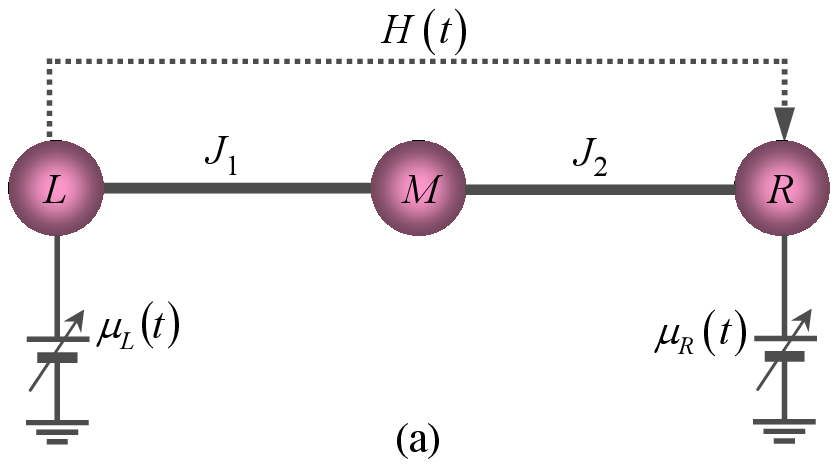} %
\includegraphics[ bb=175 336 442 555, width=6 cm, clip]{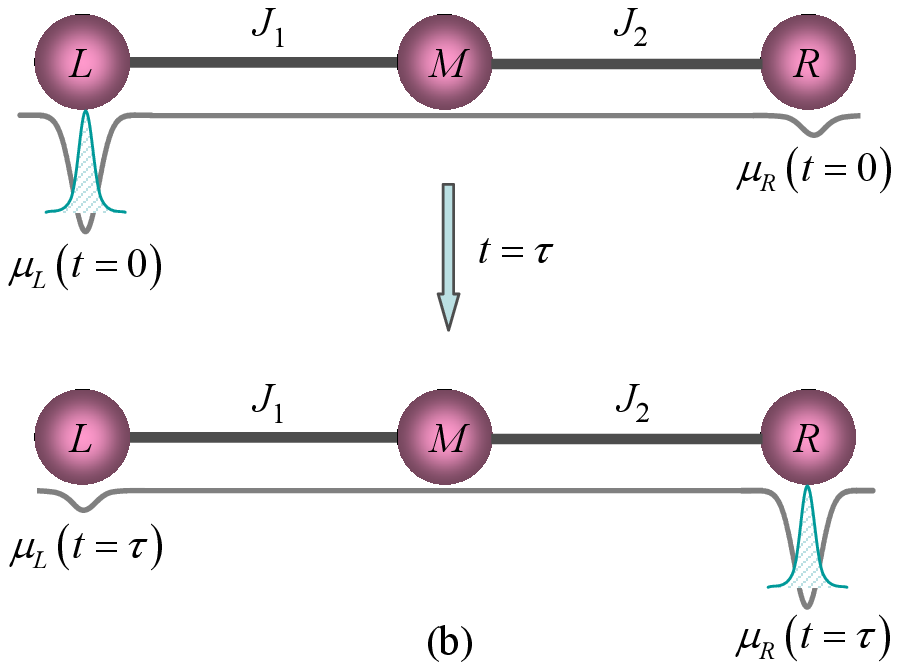}
\caption{(Color online) Schematic illustrations of quantum state transfer in
non-uniform triple-dot system: (a) the system is controlled by gates voltage
$\protect\mu _{\protect\alpha }(t)$ $(\protect\alpha =L,$ $R)$; (b) by
adiabatically change the $\protect\mu _{\protect\alpha }(t)$ $(\protect%
\alpha =L,$ $R)$ one can achieve QST from $\left\vert L\right\rangle $ to $%
\left\vert R\right\rangle $.}
\label{fig1}
\end{figure}

In this section, we first introduce the isolated (no coupling to the leads)
TQD array $\left\vert L,\sigma \right\rangle $, $\left\vert M,\sigma
\right\rangle $, $\left\vert R,\sigma \right\rangle $ ($\sigma =\uparrow
,\downarrow $), where $\left\vert L,\sigma \right\rangle =c_{L,\sigma
}^{\dag }\left\vert \text{vac}\right\rangle $ ($\left\vert M\right\rangle
=c_{M,\sigma }^{\dag }\left\vert \text{vac}\right\rangle $, $\left\vert
R\right\rangle =c_{R,\sigma }^{\dag }\left\vert \text{vac}\right\rangle $)
corresponds to an electron in the left (center, right) dot with spin $\sigma
$. The scheme is schematically shown in Fig.~\ref{fig1}(a). Specifically, we
consider the interactions between nearest-neighbor quantum dots are timeless
and slightly different. We term this model non-uniform triple-quantum-dot
(NUTQD) system. The system are controlled by external time-varying gates
voltage $\mu _{\alpha }(t)$ $(\alpha =L,$ $R)$, which control the site
energies of two end of the array. In this proposal we will show that the
information encoded in electronic spin\ can\ be transported from $\cos
\theta \left\vert L,\uparrow \right\rangle +\sin \theta \left\vert
L,\downarrow \right\rangle $ to $\cos \theta \left\vert R,\uparrow
\right\rangle +\sin \theta \left\vert R,\downarrow \right\rangle $. Notice
that the polarization of the spin of an electron is not changed as time
evolves. Then the problem about the quantum information transfer (QIT) can
be reduced to the issue of QST and a complete QST can achieve perfect QIT.
In this sense, we can ignore spin degrees of freedom to illustrate the
principles of QST from $\left\vert L\right\rangle $ to $\left\vert
R\right\rangle $.

We use $\left\{ \left\vert L\right\rangle ,\left\vert M\right\rangle
,\left\vert R\right\rangle \right\} $ as basis of the Hilbert space, the
Hamiltonian for NUTQD system in matrix form reads as
\begin{equation}
H=\left[
\begin{array}{ccc}
\mu _{L}(t) & J_{1} & 0 \\
J_{1} & 0 & J_{2} \\
0 & J_{2} & \mu _{R}(t)%
\end{array}%
\right] ,  \label{H_t}
\end{equation}%
where $J_{i}\ (i=1,2)$ is the fixed coupling constant between
nearest-neighbor dots, assumed to be real (negative) for the sake of
simplicity. The on-site energies $\mu _{L}(t)$ and $\mu _{R}(t)$ are
modulated in Gaussian pulses to realize the adiabatic transfer, according to
(shown in Fig.~2)

\begin{subequations}
\begin{align}
\mu _{L}(t)& =-\mu _{L}^{\max }\exp \left[ -\frac{1}{2}\alpha ^{2}t^{2}%
\right] ,  \label{miuL} \\
\mu _{R}(t)& =-\mu _{R}^{\max }\exp \left[ -\frac{1}{2}\alpha ^{2}\left(
t-\tau \right) ^{2}\right] ,  \label{miuR}
\end{align}%
where $\tau $ and $\alpha $ are the total adiabatic evolution time and
standard deviation of the control pulse modulating the chemical potential of
states $\left\vert L\right\rangle $ and $\left\vert R\right\rangle $. For
simplicity we set the peak voltage of each dot to be equal $\mu _{L}^{\max
}=\mu _{R}^{\max }=\mu _{0}$ and satisfy $\mu _{0}\gg \left\vert
J_{i}\right\vert $ $(i=1,2)$. We will see below that this simplicity has no
relevance to the problem.

\begin{figure}[tbp]
\includegraphics[ bb=180 175 585 440, width=7 cm, clip]{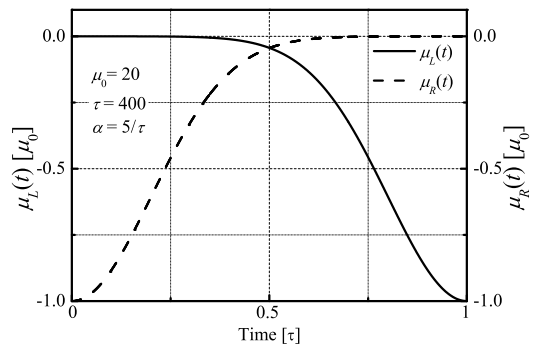}
\caption{Gate voltages as a function of time (in units of $%
\protect\tau $) , $\protect\mu _{L}(t)$ is the solid line and $\protect\mu %
_{R}(t)$ is the dash line.}
\end{figure}

At time $t=t_{0}$, the Hamiltonian $H(t_{0})$ has eigenvectors $\left\vert
\psi _{n}(t_{0})\right\rangle $ ($n=0,1,2$) which are superpositions of $%
\left\vert L\right\rangle ,$ $\left\vert M\right\rangle ,$ $\left\vert
R\right\rangle $ and the eigenvalues are denoted by $\varepsilon _{n}(t_{0})$%
, sorting in ascending order $\varepsilon _{0}<\varepsilon _{1}<\varepsilon
_{2}$. Under adiabatic evolution, these eigenstates evolve continuously to $%
\left\vert \psi _{n}(t)\right\rangle $. The instantaneous Hamiltonian's
eigen equation is

\end{subequations}
\begin{equation}
H(t)\left\vert \psi _{n}(t)\right\rangle =\varepsilon _{n}(t)\left\vert \psi
_{n}(t)\right\rangle .  \label{schrodinger}
\end{equation}

In this proposal, we use ground state $\left\vert \psi _{0}(t)\right\rangle $
of Eq.~\eqref{schrodinger} to induce population transfer from state $%
\left\vert L\right\rangle $ to $\left\vert R\right\rangle $ (see Fig.~\ref%
{fig1}(b)). One advantage of this proposal is that there can be\ no heat
dissipation during the transfer.

Starting from $t=0$, the Hamiltonian is approximate separable in the case $%
\mu _{0}\gg \left\vert J_{i}\right\vert $:
\begin{equation}
H(t=0)\simeq H_{L}\oplus H_{MR},
\end{equation}%
with
\begin{subequations}
\begin{align}
H_{L}& =-\mu _{0}\left\vert L\right\rangle \left\langle L\right\vert , \\
H_{MR}& =J_{2}\left( \left\vert M\right\rangle \left\langle R\right\vert
+\left\vert R\right\rangle \left\langle M\right\vert \right) .
\end{align}%
This Hamiltonian has the eigenstates

\end{subequations}
\begin{eqnarray}
\left\vert \psi _{\pm }\left( t=0\right) \right\rangle &=&\frac{1}{\sqrt{2}}
\left( \left\vert M\right\rangle \pm \left\vert R\right\rangle \right) ,
\notag \\
\left\vert \psi _{0}\left( t=0\right) \right\rangle &=&\left\vert
L\right\rangle ,
\end{eqnarray}%
the energies of these states are

\begin{equation}
\varepsilon _{\pm }=\pm J_{2},\text{ }\varepsilon _{0}=-\mu _{0}.
\end{equation}

Our aim is to induce population transfer from state $\left\vert
L\right\rangle $ to $\left\vert R\right\rangle $ by maintaining the system
in ground state. Now we will show that an adiabatic change of $\mu _{L}(t)$
and $\mu _{R}(t)$ will realize the QST.

In the adiabatic limit, $t\rightarrow \tau $, the parameter $\mu _{L}(t)$
goes to zero and $\mu _{R}(t)$ goes to $-\mu _{0}$. The Hamiltonian
adiabatically evolves to
\begin{equation}
H(t=\tau )\simeq H_{LM}\oplus H_{R},
\end{equation}%
with
\begin{subequations}
\begin{align}
H_{LM}& =J_{1}\left( \left\vert L\right\rangle \left\langle M\right\vert
+\left\vert M\right\rangle \left\langle L\right\vert \right) , \\
H_{R}& =-\mu _{0}\left\vert R\right\rangle \left\langle R\right\vert ,
\end{align}%
the corresponding eigenstate are
\end{subequations}
\begin{eqnarray}
\left\vert \psi _{\pm }\left( t=\tau \right) \right\rangle &=&\frac{1}{\sqrt{%
2}}\left( \left\vert L\right\rangle \pm \left\vert M\right\rangle \right) ,
\notag \\
\left\vert \psi _{0}\left( t=\tau \right) \right\rangle &=&\left\vert
R\right\rangle .
\end{eqnarray}%
and then the ground state evolves to be $\left\vert R\right\rangle $.

Providing adiabaticity is satisfied~\cite{Blum}

\begin{equation}
\left\vert \varepsilon _{m}-\varepsilon _{n}\right\vert \gg |\langle \psi
_{m}|\dot{\psi}_{n}\rangle |,  \label{AC}
\end{equation}%
the overall system will remain in its instantaneous ground state. At $t=0$,
the system is prepared in state $\left\vert \psi _{0}\left( t=0\right)
\right\rangle =\left\vert L\right\rangle $, then the adiabatic theorem
states that the system will stay in $\left\vert \psi _{0}\left( t\right)
\right\rangle $. Note that $\left\vert L\right\rangle $ and $\left\vert
R\right\rangle $ denote the states in which the electron is on the left and
right QD, respectively. Therefore, we can see that an electron starting in $%
\left\vert L\right\rangle $ will end up in $\left\vert R\right\rangle $.

Providing the length of time $\tau $ is too large, that is, the
time-dependent change is introduced slowly enough, the fidelity of QST is
also determined by peak gate voltage $\mu _{0}$. Notice that the square of
the module of fidelity $\left\vert F(t)\right\vert ^{2}=\left\vert
\left\langle R\right. \left\vert \psi _{0}\left( t\right) \right\rangle
\right\vert ^{2}$ denotes the probability of finding $\left\vert
R\right\rangle $ in the ground state $\left\vert \psi _{0}\left( t\right)
\right\rangle $. Now we suppose to get analytical expression of fidelity
using first order perturbation theory. We start from Eq.~\eqref{H_t} at $%
t=\tau $ and consider the coupling term $J_{2}\left( \left\vert
R\right\rangle \left\langle M\right\vert +\left\vert M\right\rangle
\left\langle R\right\vert \right) $ as a weak perturbation. The Hamiltonian

\begin{equation}
H(t=\tau )=H_{0}+H_{I},
\end{equation}%
contains two parts
\begin{subequations}
\begin{align}
H_{0}& =J_{1}\left( \left\vert L\right\rangle \left\langle M\right\vert
+\left\vert M\right\rangle \left\langle L\right\vert \right) -\mu
_{0}\left\vert R\right\rangle \left\langle R\right\vert , \\
H_{I}& =J_{2}\left( \left\vert R\right\rangle \left\langle M\right\vert
+\left\vert M\right\rangle \left\langle R\right\vert \right) .
\end{align}

Our aim is to find the approximate expression for the ground state $%
\left\vert \psi _{0}\right\rangle $ of the perturbed Hamiltonian $H(t=\tau )$%
. The eigenfunctions of unperturbed Hamiltonian $H_{0}$ is
\end{subequations}
\begin{eqnarray}
|\psi _{0}^{(0)}\rangle &=&|R\rangle ,  \notag \\
|\psi _{\pm }^{(0)}\rangle &=&\frac{1}{\sqrt{2}}\left( \left\vert
L\right\rangle \pm \left\vert M\right\rangle \right) .
\end{eqnarray}%
In the picture of $\left\{ |\psi _{-}^{(0)}\rangle ,|\psi _{+}^{(0)}\rangle
,|\psi _{0}^{(0)}\rangle \right\} $, The Hamiltonian $H_{0}$ can be
diagonalized as
\begin{equation*}
H_{0}=\left[
\begin{array}{ccc}
-J_{1} & 0 & 0 \\
0 & J_{1} & 0 \\
0 & 0 & -\mu _{0}%
\end{array}%
\right] .
\end{equation*}

As the first order perturbation, we have the corrected ground state to be%
\begin{eqnarray}
\left\vert \psi _{0}\right\rangle &=&|\psi _{0}^{(0)}\rangle +\sum_{\eta
=\pm }\frac{\langle \psi _{\eta }^{(0)}|H_{I}|\psi _{0}^{(0)}\rangle }{%
E_{0}^{(0)}-E_{\eta }^{(0)}}|\psi _{\eta }^{(0)}\rangle  \notag \\
&=&\frac{J_{1}J_{2}}{\mu _{0}^{2}-J_{1}^{2}}|L\rangle -\frac{\mu _{0}J_{2}}{%
\mu _{0}^{2}-J_{1}^{2}}|M\rangle +|R\rangle .
\end{eqnarray}%
So the transfer fidelity of adiabatic QST at $t=\tau $\ is

\begin{eqnarray}
\left\vert F(\tau )\right\vert ^{-2} &=&1+\left( \frac{J_{1}J_{2}}{\mu
_{0}^{2}-J_{1}^{2}}\right) ^{2}+\left( \frac{\mu _{0}J_{2}}{\mu
_{0}^{2}-J_{1}^{2}}\right) ^{2}  \notag \\
&=&1+\frac{J_{2}^{2}\left( \mu _{0}^{2}+J_{1}^{2}\right) }{\left( \mu
_{0}^{2}-J_{1}^{2}\right) ^{2}},  \label{Fidelity}
\end{eqnarray}%
which shows that the peak voltage $\mu _{0}$ determined the fidelity of QST.
As $\mu _{0}\gg \left\vert J_{i}\right\vert $\ is satisfied, the fidelity is
near to unity.

\section{Numerical Simulations}

The analysis above is based on the assumption that the adiabaticity is
satisfied. In order to demonstrate the QST in the system~\eqref{H_t} and to
show how exact the approximation is, in this section we numerically solve
the master equation and the above central conclusion can be get confirmed.
The main goal of this section is to analyze the parameters which influence
the fidelity of adiabatic QST and find the proper matching relation between
them.

First, initialize electron in the left dot, i.e., the total initial state is
$\left\vert \Psi \left( 0\right) \right\rangle =\left\vert L\right\rangle $,
the time evolution creates a coherent superposition:
\begin{equation}
\left\vert \Psi \left( t\right) \right\rangle =c_{1}(t)\left\vert
L\right\rangle +c_{2}(t)\left\vert M\right\rangle +c_{3}(t)\left\vert
R\right\rangle .
\end{equation}%
with this notation we assume the initial condition $c_{1}(0)=1$, and the
other two equal zero. In order to proceed, we numerically solve the master
equations for the density matrix $\rho $. The master equation is written as~%
\cite{Blum} (assuming $\hbar =1$)

\begin{equation}
i\frac{d\rho \left( t\right) }{dt}=\left[ H,\rho \left( t\right) \right] ,
\end{equation}%
where $\rho \left( t\right) =\left\vert \Psi \left( t\right) \right\rangle
\left\langle \Psi \left( t\right) \right\vert $. With the basis state
ordering $\left\{ \left\vert L\right\rangle ,\left\vert M\right\rangle
,\left\vert R\right\rangle \right\} $, the density matrix can be written as
\begin{equation*}
\rho \left( t\right) =\left[
\begin{array}{ccc}
\left\vert c_{1}(t)\right\vert ^{2} & c_{1}(t)c_{2}^{\ast }(t) &
c_{1}(t)c_{3}^{\ast }(t) \\
c_{2}(t)c_{1}^{\ast }(t) & \left\vert c_{2}(t)\right\vert ^{2} &
c_{2}(t)c_{3}^{\ast }(t) \\
c_{3}(t)c_{1}^{\ast }(t) & c_{3}(t)c_{2}^{\ast }(t) & \left\vert
c_{3}(t)\right\vert ^{2}%
\end{array}%
\right] \;.
\end{equation*}

\begin{figure}[tbp]
\includegraphics[ bb=86 84 290 368, width=7 cm, clip]{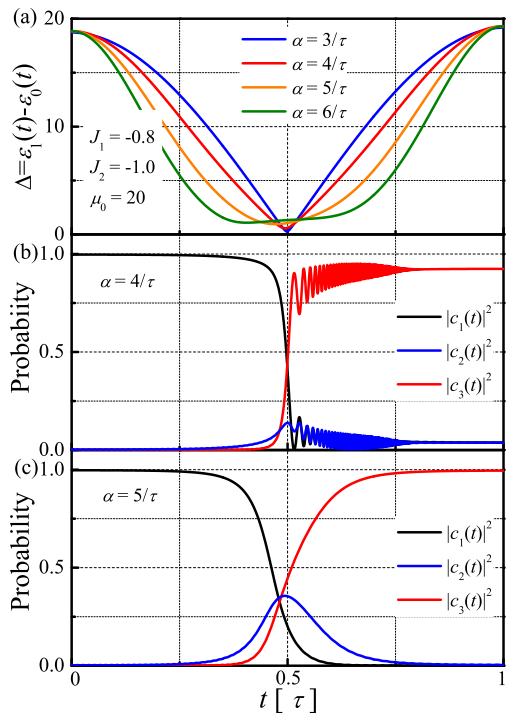}
\caption{(Color online) (a) The energy gap $\Delta (t)=\protect\varepsilon %
_{1}(t)-\protect\varepsilon _{0}(t)$ between the first-excited state and
ground state of the triple-dot system undergoing evolution due to modulation
of the gate volgate according to pulse Eq. (3) for $\protect\mu _{0}=20$, $%
J_{1}=0.8$, $J_{2}=1.0$, $\protect\tau =10\protect\mu _{0}/J_{1}^{2}$ and $%
\protect\alpha =3/\protect\tau $, $4/\protect\tau $, $5/\protect\tau $, $6/%
\protect\tau $. The time evolution of the probabilities induced by the
pulses in Fig. 2 for (b) $\protect\alpha =4/\protect\tau $ and (c) $\protect%
\alpha =5/\protect\tau $. Initially the population is on left qubit (black
line) and finally mainly on right qubit (red line). The population on the
intermediate qubit is shown as a blue line.}
\label{fig3}
\end{figure}

According to the definition of fidelity, we can see that $\left\vert
F(t)\right\vert ^{2}=\left\vert c_{3}(t)\right\vert ^{2}$. The crucial
requirement for adiabatic evolution is Eq.~\eqref{AC}. Firstly, one must to
make sure that no level crossings occur, i.e., $\varepsilon
_{0}(t)-\varepsilon _{j}(t)<0$. To calculate the energies is generally only
possible numerically. In Fig.~\ref{fig3}(a) we present the results showing
the eigenenergy gap $\Delta (t)=\varepsilon _{1}(t)-\varepsilon _{0}(t)$
between the first-excited state and ground state of the NUTQD system
undergoing evolution due to modulation of the gate voltage according to
pulse Eq.~\eqref{H_t} for $\mu _{0}=20$, $J_{1}=0.8$, $J_{2}=1.0$, $\tau
=10\mu _{0}/J_{1}^{2}$ and $\alpha =3/\tau $, $4/\tau $, $5/\tau $, $6/\tau $%
. It shows that for the given evolution time $\tau =400$ the minimum of the
energy gap decrease as standard deviation $\alpha $ increasing. The slower
Hamiltonian~\eqref{H_t} varies, the closer adiabatic theorem holds. In Fig.~%
\ref{fig3} we also show the numerically computed behavior of the populations
$|c_{1}(t)|^{2}$, $|c_{2}(t)|^{2}$ and $|c_{3}(t)|^{2}$ on the three quantum
dots as a function of time with $\alpha =4/\tau $ and $\alpha =5/\tau $.
Note that for $\alpha =4/\tau $ transfer, as illustrated in Fig.~\ref{fig3}%
(b), the population on state $\left\vert R\right\rangle $ is decoupled and
stays constant 0.92. The fraction of population left in states $\left\vert
L\right\rangle $ and $\left\vert M\right\rangle $ is $\left\vert c_{1}(\tau
)\right\vert ^{2}+\left\vert c_{2}(\tau )\right\vert ^{2}=0.08$ and executes
Rabi oscillations because the quantum dots $L$ and $M$ are coupled with $%
J_{1}=0.8$. Whereas for $\alpha =5/\tau $ case, shown in Fig.~\ref{fig3}(c),
one can see that the fidelity of adiabatic QST has been improved
considerably by this slight change. The fidelity of QST achieve 0.995 and
only 0.5\% of population remains in states $\left\vert L\right\rangle $ and $%
\left\vert M\right\rangle $. This is consistent with the results shown in
Fig.~\ref{fig3}(a) because the eigenenergy gap plays opposite role for
transition probability.

\begin{figure}[tbp]
\includegraphics[ bb=11 18 490 408, width=7 cm, clip]{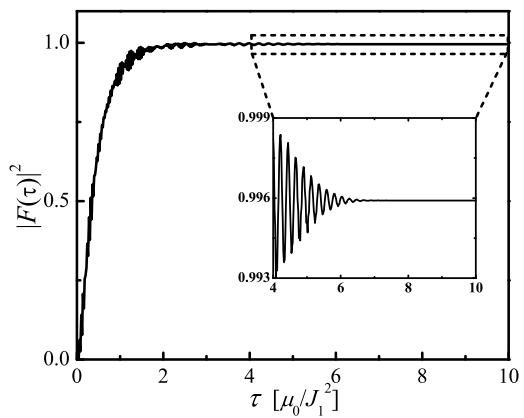}
\caption{Fidelity as a function of total adiabatic evolution time $\protect%
\tau$ (in units of $\protect\mu_{0}/J_{1}^{2}$). When $\protect\tau \geq 4%
\protect\mu_{0}/J_{1}^{2}$, the fidelity of QST becomes stable.}
\label{fig4}
\end{figure}

The fidelity of population transfer will be very high as long as the
Hamiltonian evolves sufficiently slowly in time (as determined by criteria
for the applicability of the theorem). In practice the maximum possible
transfer rates will be a few times greater than $\mu_{0}/J_{1}^{2}$ which is
illustrated in Fig.~\ref{fig4}. Note that the transfer fidelity becomes
stable when the total evolution time satisfy $\tau \geq 4\mu_{0}/J_{1}^{2}$.

\begin{figure}[ptb]
\begin{centering}
\includegraphics[ bb=162 160 508 418, width=7 cm, clip]{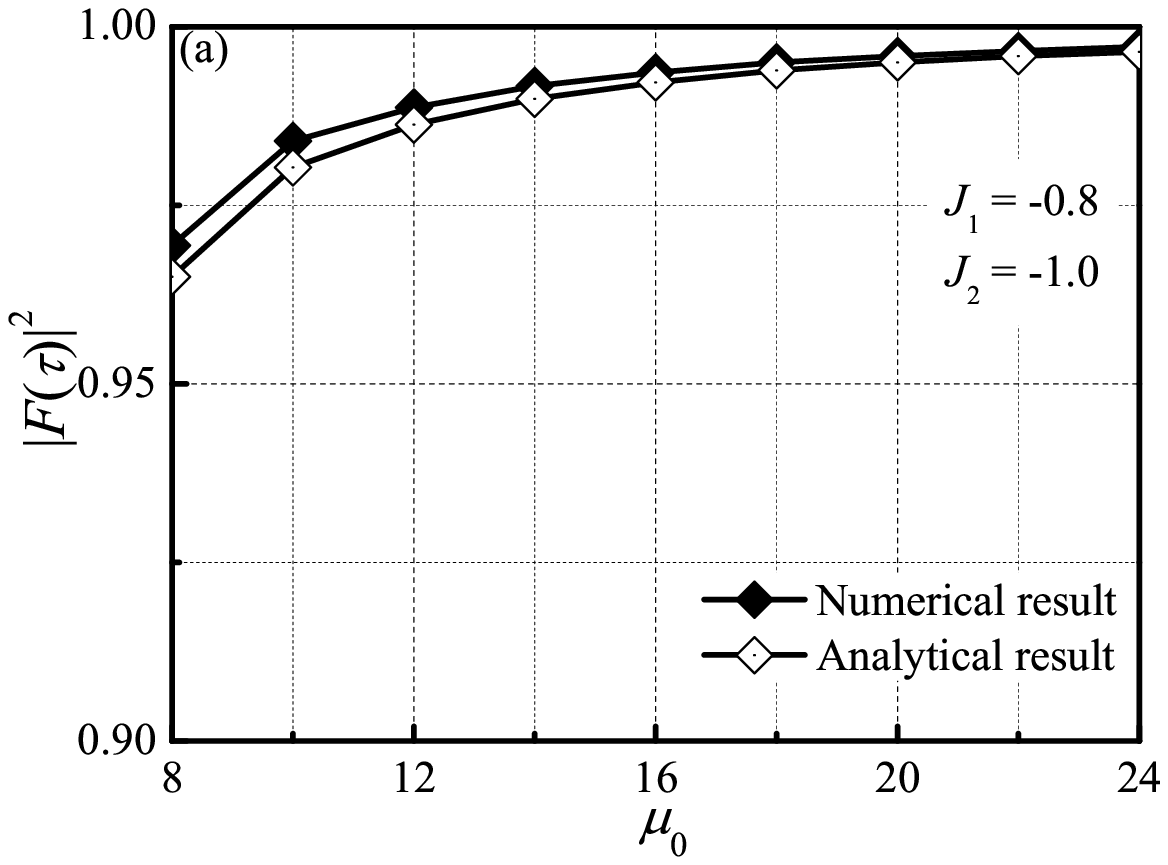} %
\includegraphics[ bb=33 38 517 400, width=7 cm, clip]{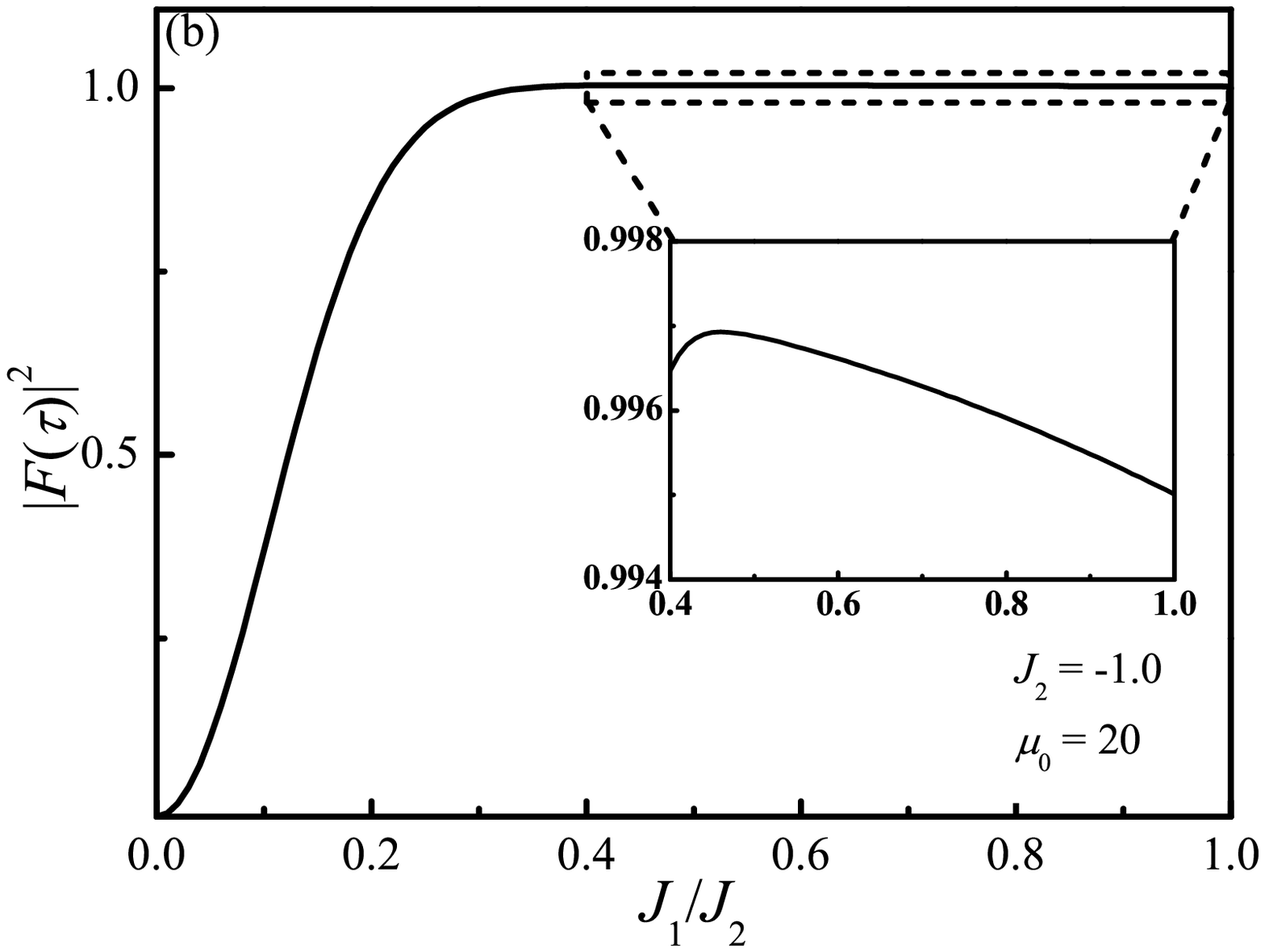}
\par\end{centering}
\caption{The plot of the square of fidelity $\left\vert F(\protect\tau%
)\right\vert ^{2}$ as a function of system paremeters: (a) the peak voltage $%
\protect\mu _{0}$ and (b) the ratio $J_{1}/J_{2}$. If the condition is
satisfied when $|\protect\mu _{0}/J_{max}|\geq 14$ and $J_{1}/J_{2}\geq 0.4$%
, the transfer fidelity is near to one. }
\label{fig5}
\end{figure}

The preceding discussion is based on the assumption that the system
parameters are setup with arbitrary precision that is the system is coupled
with $J_{1}=0.8$ and $J_{2}=1.0$. However, it is difficult to fabricate such
precise Hamiltonian in experiment. Next we will show that the adiabatic
passage like us is relatively insensitive to the system parameters. From the
analytical results, the fidelity of adiabatic QST depends on the contrast
ratio between peak voltage $\mu _{0}$ and coupling constants $J_{i}$. To
determine the parameter range needed to achieve high fidelity transfer, we
numerically integrate the density matrix equations of motion, with varying
the peak voltage $\mu _{0}$. In Fig.~\ref{fig5}(a) we present results
showing the square of fidelity $\left\vert F(\tau )\right\vert
^{2}=|c_{3}(\tau )|^{2}$ as a function of $\mu _{0}$ with $J_{1}=0.8$, $%
J_{2}=1.0$, $\tau =375$ and $\alpha =5/\tau $. We can see that the
population transfer is close to one ($\left\vert F(\tau )\right\vert
^{2}\geq 0.99$) and stable when $\mu _{0}$ is achieved for $|\mu
_{0}/J_{2}|\geq 14$. The plot in Fig.~\ref{fig5}(a) is in agreement with the
analytical results Eq.~\eqref{Fidelity} with high accuracy. On the other
hand, the difference between $J_{1}$ and $J_{2}$ has a little effect upon
transfer fidelity within certain range. We have illustrated this in Fig.~\ref%
{fig5}(b) where the effects of mismatch between $J_{1}\ $and $J_{2}$ have
been modeled. Here we show $\left\vert F(\tau )\right\vert ^{2}$ as a
function of $J_{1}/J_{2}$ for peak voltage $\mu _{0}=20$ to simulate the
effect of a systematic error in the coupling constants. Note that the ratio
as much as $0.35$ still permits $\left\vert F(\tau )\right\vert ^{2}\approx
0.994$.

\section{Summary and discussion}

In summary, we have introduced a method of coherent QST through a NUTQD
system by adiabatic passage. This scheme is realized by modulation of gate
voltage of QDs. Different from the CTAP Scheme, our method is to induce
population transfer by maintaining the system in its ground state which is
more stable than dark state. We have studied the adiabatic QST through a
NTQD system by theoretical analysis and numerical simulations of the ground
state evolution of NTQD model. The result shows that it is a high fidelity
process for a proper choose of standard deviation and peak voltage.

In order to investigate the relation between the fidelity of quantum state
transfer $\left\vert F(\tau )\right\vert ^{2}$ and peak voltage $\mu _{0}$,
we have numerically solve the master equation under different peak voltage.
The numerical result shows that if we want to achieve a high fidelity more
than 99.5\% we require the ratio of $\left\vert \mu _{0}/J_{2}\right\vert
\geq 14$. We also show that the sight difference between $J_{1}$ and $J_{2}$
does small influence on the fidelity.

It is worthwhile to discuss the applicability of the scheme presented above.
In a real system, quantum decoherence is the main obstacle to the
experimental implementation of quantum information. For coupled QDs,
experiments \cite{T1} show that the coupling strength $J$ is about 0.25 meV
while $\mu _{0}\sim 20J$. we can estimate a time of $\sim $ 50 ps required
for adiabatic operation. On the other hand, the typical decoherence time $%
T_{2}$ for electron-spin has been indicated experimentally \cite{T2} to be
longer than 80$\pm $9 $\mu $s at 2.5 K which is much longer than adiabatic
operation time. So our scheme has applicability in practice.

\begin{acknowledgements}
One of the authors (BC) thanks Z. Song for discussions and encouraging comments.
We also acknowledge the support of the NSF of China (Grant Nos. 10847150, 61071016) and Shandong Provincial Natural
Science Foundation, China (Grant No. ZR2009AM026).
\end{acknowledgements}

\end{document}